\documentclass{ws-p8-50x6-00}

\def\ifmath#1{\relax\ifmmode #1\else $#1$\fi}%
\def\rd{\ifmath{{\mathrm{d}}}}

\def\stat{\ifmath{{\mathrm{stat}}}}
\def\syst{\ifmath{{\mathrm{syst}}}}
\def\stat{\ifmath{{\mathrm{stat}}}}

\def\GeV{\ifmath{{\mathrm{Ge\hspace{-.6mm}V}}}}
\def\MeV{\ifmath{{\mathrm{Me\hspace{-.6mm}V}}}}

\def\rand{\ifmath{{\mathrm{and}}}}

\def\re{\ifmath{{\mathrm{e}}}}

\def\rW{\ifmath{{\mathrm{W}}}}
\def\rq{\ifmath{{\mathrm{q}}}}

\def\pairs{\ifmath{{\mathrm{pairs}}}}
\def\ev{\ifmath{{\mathrm{ev}}}}

\def\mix{\ifmath{{\mathrm{mix}}}}
\def\MC{\ifmath{{\mathrm{MC}}}}
\def\noBE{\ifmath{{\mathrm{noBE}}}}

\def\pairs{\ifmath{{\mathrm{pairs}}}}

\def\exp{\ifmath{{\mathrm{exp}}}}

\begin{document}

\title{Bose-Einstein Correlations in WW Events at LEP2}

\author{J.A. van Dalen}

\address{HEFIN, University of Nijmegen/NIKHEF,
Toernooiveld 1, 6525 ED Nijmegen, NL\\
E-mail: jornvd@hef.kun.nl}


\maketitle

\abstracts{
Analyses of Bose-Einstein Correlations in $\rW^{+}\rW^{-}$ events at LEP2 by the four
LEP collaborations are presented. In particular, Bose-Einstein correlations in
$\rW^{+}\rW^{-}$ overlap are investigated and the possible existence of these 
correlations between particles coming from different W's, which may influence the
W-mass measurements in the fully hadronic channel $\re^{+}\re^{-}\rightarrow
\rW^{+}\rW^{-}\rightarrow\rq_1 \bar{\rq}_2 \rq_3 \bar{\rq}_4 $. No evidence for
such an inter-W Bose-Einstein correlation is found. The results are preliminary.
}

\section{Introduction}
Bose-Einstein interference is observed in hadronic Z-decay as an
enhanced production of identical bosons at small
four-momentum difference~\cite{kittel}. There is no reason why such an interference
should not be present within hadronic W-decay (intra-W BE interference), as well.
Indeed, the four LEP collaborations measure intra-W Bose-Einstein correlations (BEC),
as a consequence of intra-W BE interference, see~\cite{delphiea} and references therein.
Furthermore, since in fully hadronic WW events
($\rW^{+}\rW^{-}\rightarrow \rq_{1}\bar{\rq}_{2}\rq_{3}\bar{\rq}_{4}$)
the W-decay products
overlap in space-time at LEP2 energies, it may be natural
to expect interference also between identical bosons originating from different
W's (inter-W BE interference). Together with colour reconnection~\cite{abreu},
inter-W BE interference not only forms a potential bias in the determination
of the W-mass, but also may provide a laboratory to measure the space-time
development of this overlap. 
In this contribution, the latest preliminary results of the four LEP experiments on inter-W BEC
are presented.~\cite{delphiea}

\section{Analysis}\label{sect2}
The largely model independent method used to study inter-W BEC is based on that proposed in.~\cite{cwk}
If the two W's would decay independently, the two-particle density in fully hadronic WW events, $\rho_{2}^{\rW\rW}$,
would be given by
\begin{eqnarray}
  \rho_{2}^{\rW\rW}\!(p_1,p_2)\! &=& \rho_{2}^{\rW^{+}}\!(p_1,p_2) \!+\! \rho_{2}^{\rW^{-}}\!(p_1,p_2)
	 \!+\! \rho_{1}^{\rW^{+}}\!(p_{1})\rho_{1}^{\rW^{-}}\!(p_{2})
	 \!+\! \rho_{1}^{\rW^{-}}\!(p_{1})\rho_{1}^{\rW^{+}}\!(p_{2})\phantom{1}     \label{eq1a}\nonumber  \\
	&=& 2\rho_{2}^{\rW}\!(p_{1},p_{2})
         \!+\! 2\rho_{1}^{\rW}\!(p_{1})\rho_{1}^{\rW}\!(p_{2})       \label{eq1b}
	 \ \ \ ,
\end{eqnarray}
where the superscripts, e.g., WW, indicate which W's decay hadronically. 
Experimentally, $\rho_{2}^{\rW\rW}$ is measured in the fully hadronic WW events
and $\rho_{2}^{\rW}$ in the semi-hadronic events, i.e., events where one W decays hadronically and
the other W decays into a charged lepton and its corresponding neutrino.
To measure the product of the single-particle densities, a two-particle
density, $\rho (p_{1},p_{2})^{\rW^{+}\rW^{-}}_{\mix}$, is constructed by pairing
particles originating from two different semi-hadronic WW events. 
By construction these pairs of particles are uncorrelated.

Since BEC are large at small values of the four-momentum difference, $Q$, between 
two identical particles, and since the statistics are too poor 
(about 10k WW events are selected per experiment at LEP2, at $160<\sqrt{s}<209\,\GeV$)
to do a multi-dimensional analysis, 
the two-particle densities are measured in this one-dimensional distance measure:
$\rho_2 (Q)\equiv \frac{1}{N_{\ev}}\frac{\rd n_{\pairs}}{\rd Q}$,
where $N_{\ev}$ is the number of events that is used
and $n_{\pairs}$ is the number of particle pairs.

Deviations from the hypothesis that the two W's decay independently can be found in deviations from
Eq.\,(\ref{eq1b}). Therefore, the following test statistics are used:
\begin{eqnarray}
\Delta\rho\equiv\rho_{2}^{\rW\rW}-2\rho_{2}^{\rW}-2\rho^{\rW^{+}\rW^{-}}_{\mix} \ \ \ \rand \ \ \ 
D\equiv\frac{\rho_{2}^{\rW\rW}}{2\rho_{2}^{\rW}+2\rho^{\rW^{+}\rW^{-}}_{\mix}} \ \ \ . \label{eq3}
\end{eqnarray}
To diminish artificial distortions due to event mixing, non-BEC and various detector effects,
also the double ratio $D'\equiv D/D_{\MC,\,\noBE}$ is used, where $D_{\MC,\,\noBE}$ is a MC sample
without BEC, or at least without inter-W BEC.
Evidently, $\Delta\rho\neq 0$ and $D=D'\neq 1$ in the case of inter-W BEC.

\section{Results}
\begin{figure}

\vspace{-.6cm}
\begin{tabular}{cc}

\hspace{-.8cm}
\epsfig{figure=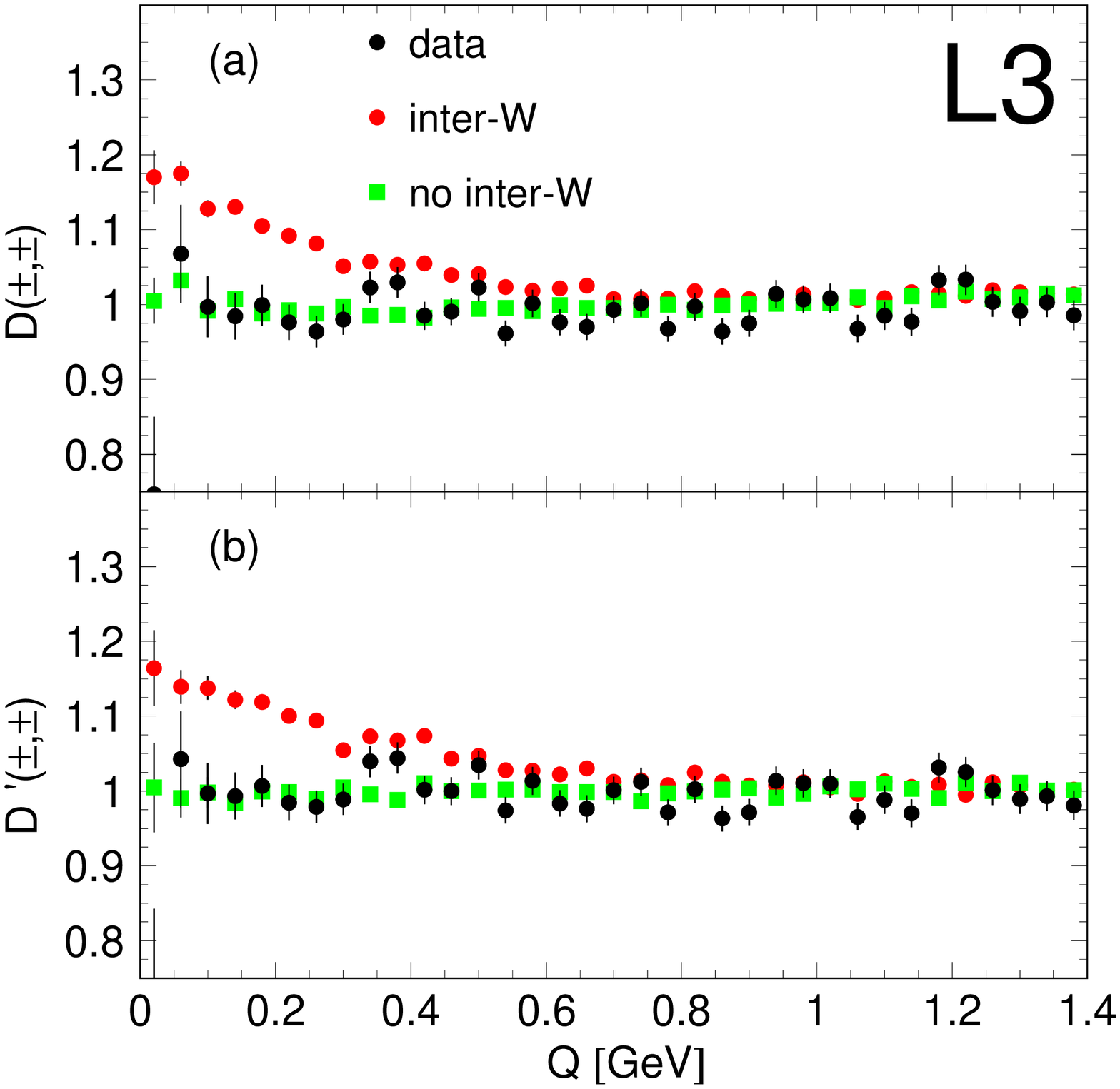, width=.55\linewidth} &

\vspace{-.4cm}
\hspace{-.7cm}
\epsfig{figure=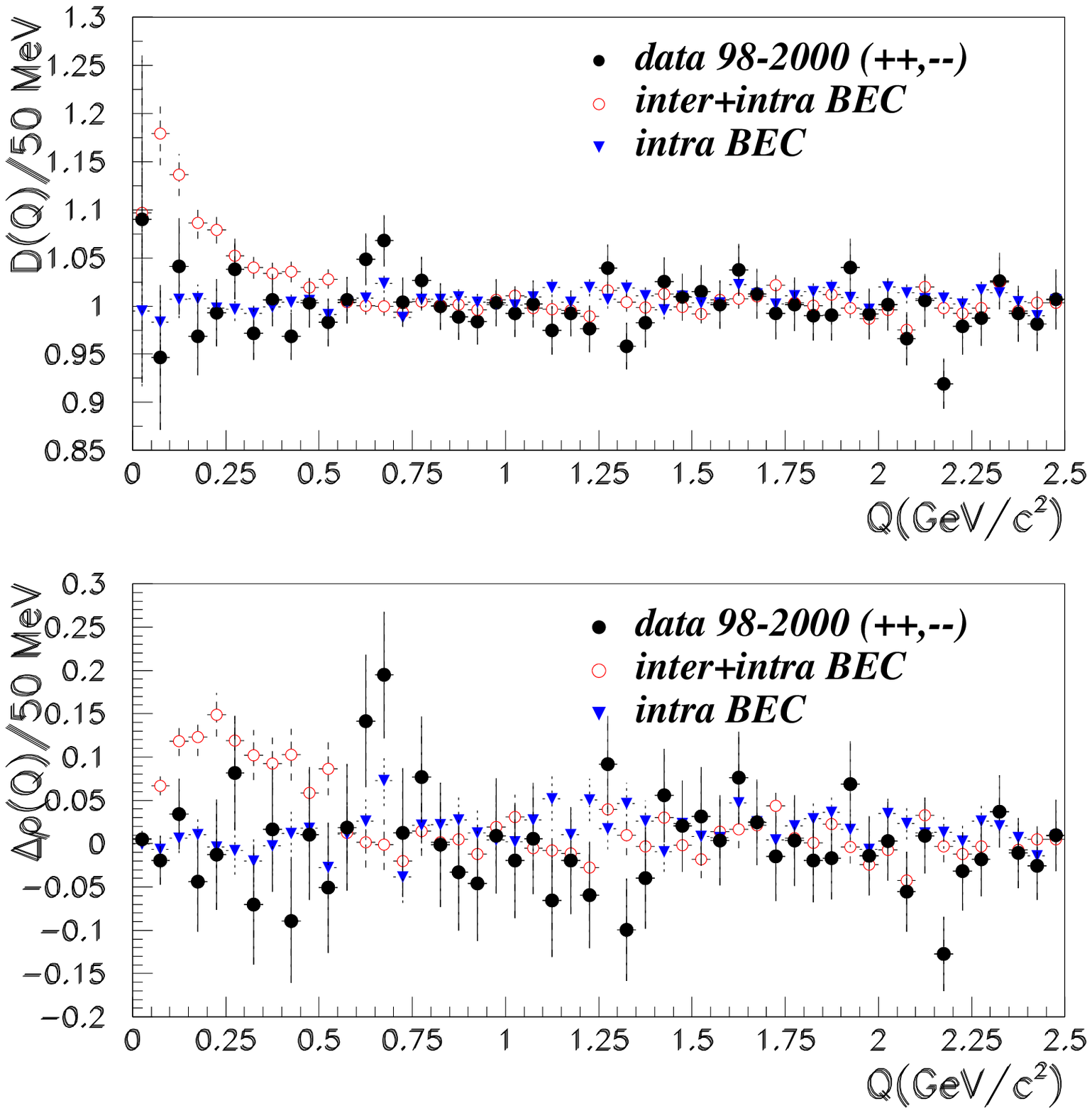, height=7.0cm, width=.55\linewidth}
\end{tabular}
\begin{minipage}[t]{.46\linewidth}
 \vspace{-.15cm}
 \caption{The ratios $D$, (a), and $D'$, (b), for like-sign pairs,
 using the L3 data.}
\label{fig1}
\end{minipage}
\hspace{.8cm}
\begin{minipage}[t]{.46\linewidth}
 \vspace{-.15cm}
 \caption{The test statistics $D(Q)$ and $\Delta\rho(Q)$, using
 the DELPHI data.}
\label{fig2}
\end{minipage}

\vspace{-.7cm}
\end{figure}

Fig.~\ref{fig1} shows the distributions of $D$ and $D'$ for like-sign track pairs, for the L3 data collected
in the period 1998-2000. It is clear that the data show flat distributions in both cases (the same conclusion
is obtained for the $\Delta\rho(Q)$ distribution, which is not shown), indicating no
evidence for inter-W BEC and $D\simeq D'$ (i.e., only a small model dependence). Also shown in the figure are the predictions
of a MC with the BE$_{32}$ algorithm~\cite{be32} and tuned to the Z-data, 
for the scenario where correlations
between all particles are allowed (BEA) and for the scenario where only intra-W BEC are allowed (BES).
It is clear that the BEA scenario is disfavoured, while BES describes the data. In particular, $D'(Q)$ is fitted
by $(1+\varepsilon Q)(1+\Lambda \exp(-k^2 Q^2))$,
where $\varepsilon$, $\Lambda$ and $k$ are the fit parameters. The parameter $\Lambda$
measures the strength of inter-W BEC. The result of the fit is $\Lambda=0.008\pm0.018(\stat)\pm0.016(\syst)$.
This can be interpreted as evidence against BEC between identical particles originating from
different W's.

Analogously, Fig.~\ref{fig2} shows $D(Q)$ and $\Delta\rho(Q)$ for the \mbox{DELPHI} data.
The same conclusions are obtained as for the L3 data:
there is no evidence for inter-W BEC and a MC with inter-W BEC is clearly disfavoured. In particular,
$D(Q)$ is fitted by $\gamma(1+\varepsilon Q)(1+\Lambda \exp(-k Q))$. Fixing $k$ at 1.01 fm (computed from
a MC with inter-W BEC, using BE$_{32}$), one obtained $\Lambda=-0.037\pm 0.055(\stat)\pm 0.055(\syst)$,
consistent with no inter-W BEC.

The ALEPH collaboration also computed the $\Delta\rho(Q)$ and $D(Q)$ test statistics. Qualitatively, the
same observations are made as by L3 and DELPHI: the data show flat distributions, indicating no evidence
for inter-W BEC. A MC with BEA is disfavoured and the BES scenario describes the data. No numbers
are computed, yet. The `standard' analysis of ALEPH differs from the method explained in Sect.~\ref{sect2}. 
It starts by defining a correlation function $R_2 \equiv \rho_2 / \rho_0$, where $\rho_0$ is the two-particle
density that would occur in a world without BEC. This correlation function is computed for the 
selected fully hadronic WW events and for two MC scenarios, BEA and BES, using the BE$_{3}$ algorithm~\cite{be32}
(the MC is tuned on a light-quark Z-decay
sample). One observes that the BES scenario describes the data, whereas BEA is disfavoured. Although
this result agrees with the results from the method explained in Sect.~\ref{sect2}, it should be
realized that there is a strong model dependence, due to the MC that is needed to compute $\rho_0$, and
that $R_2$ is also sensitive to intra-W BEC. These are two disadvantages with respect to the method of
Sect.~\ref{sect2}.

Unfortunately, the OPAL collaboration has no new results since 1999. A new analysis, using
the method of Sect.~\ref{sect2} is in progress, but is not finalized, yet. 
The results of the `old' analysis are based on a completely different method and due to the lack of
statistics the results are inconclusive, but not in contradiction with the others.

\section{Summary and Interpretation of the Results}

A common method is used by L3, DELPHI and ALEPH (and will be used by OPAL), to search for
inter-W BEC. No evidence is found for the existence of these correlations and MC models allowing
inter-W BEC are (greatly) disfavoured. The picture is now  more consistent than a year ago:~\cite{sharkaismd}
There is a generally accepted common method and the results from all experiments
are more or less consistent. This provides a basis to
combine results of the four LEP experiments into one single number in the near future.

Provided that there is no colour reconnection, the Lund string model~\cite{bo} can only
give the BE effect between bosons coming from the same string.
Thus, the data, that do not show any indication for inter-W BEC, are no problem for this model.

The positive outcome of the BE analyses, is that the effect on the W mass measurement
due to inter-W BEC, is small. At the time that W production started
at LEP2, the worst case scenario predicted an effect on the measurement of the W mass up to 
100 $\MeV$.~\cite{losj} Before the results shown in this contribution, 
the L3 collaboration estimated the error to be 60
$\MeV$.~\cite{l3mw1} Now, it has been reduced to 20 $\MeV$.~\cite{l3mw2}
The error is expected to go down even further, when the
LEP results are combined.

\end{document}
